# Point-contact Andreev-reflection spectroscopy in ReFeAsO$_{1-x}$F$_x$ (Re = La, Sm): Possible evidence for two nodeless gaps

R. S. Gonnelli[1], D. Daghero[1], M. Tortello[1], G. A. Ummarino[1], V. A. Stepanov[2], R. K. Kremer[3], J. S. Kim[3], N. D. Zhigadlo[4], and J. Karpinski[4]

[1] *Dipartimento di Fisica and CNISM, Politecnico di Torino, corso Duca degli Abruzzi 24, 10129 Torino (TO) – Italy*
[2] *P.N. Lebedev Physical Institute, Russian Academy of Sciences, Leninskiy Prospekt 53, 119991 Moscow, Russia*
[3] *Max-Planck-Institut für Festkörperforschung, D-70569 Stuttgart, Germany*
[4] *Laboratory for Solid State Physics, ETHZ, CH-8093 Zurich, Switzerland*

**Abstract**

A deep understanding of the character of superconductivity in the recently discovered Fe-based oxypnictides ReFeAsO$_{1-x}$F$_x$ (Re = rare-earth) necessarily requires the determination of the number of the gaps and their symmetry in k space, which are fundamental ingredients of any model for the pairing mechanism in these new superconductors. In the present paper, we show that point-contact Andreev-reflection experiments performed on LaFeAsO$_{1-x}$F$_x$ (La-1111) polycrystals with T$_c$ ~ 27 K and SmFeAsO$_{0.8}$F$_{0.2}$ (Sm-1111) polycrystals with T$_c$ ~ 53 K gave differential conductance curves exhibiting two peaks at low bias and two additional structures (peaks or shoulders) at higher bias voltages, an experimental situation quite similar to that observed by the same technique in pure and doped MgB$_2$. The single-band Blonder-Tinkham-Klapwijk model is totally unable to properly fit the conductance curves, while the two-gap one accounts remarkably well for the shape of the whole experimental dI/dV vs. V curves. These results give direct evidence of two nodeless gaps in the superconducting state of ReFeAsO$_{1-x}$F$_x$ (Re = La, Sm): a small gap, $\Delta_1$, smaller than the BCS value (2$\Delta_1$ / k$_B$T$_c$ ~ 2.2 – 3.2) and a much larger gap $\Delta_2$ which gives a ratio 2$\Delta_2$ / k$_B$T$_c$ ~ 6.5 – 9. In Sm-1111 both gaps close at the same temperature, very similar to the bulk T$_c$, and follow a BCS-like behaviour, while in La-1111 the situation is more complex, the temperature dependence of the gaps showing remarkable deviations from the BCS behaviour at T close to T$_c$. The normal-state conductance reproducibly shows an unusual, but different, shape in La-1111 and Sm-1111 with a depression or a hump at zero bias, respectively. These structures survive in the normal state up to T$^*$ ~ 140 K, close to the temperatures at which structural and magnetic transitions occur in the parent, undoped compound.

## 1. Introduction

Twenty years of research about the high-T$_c$ cuprates have led to a tremendous refinement and development of theoretical and experimental techniques for the detection and investigation of even the tiniest features of these compounds. The huge expertise the scientific community has acquired in this way is now being applied to the newly discovered Fe-based superconductors [1]. However, the complexity of these compounds is considerable and the explanation of their behaviour is far from being achieved. The determination of the number, the magnitude and the symmetry of the gap(s) is of central importance in establishing the microscopic origin of superconductivity, and is the point on which many groups have focused their attention. However, the results are often inconsistent with one another, due to various factors such as the quality of the samples, the (not clear yet) generality of some features across the various classes of Fe-based compounds ("1111" versus "122" family, for example), the specific pitfalls of the different experimental techniques.

The similarity between Fe-based superconductors and cuprates, especially highlighted at the beginning of the research (but questioned by new, more accurate results) was also supported by some experimental data that were interpreted as pointing towards a (single or multiple) *d*-wave order parameter, or, at least, an unconventional order parameter with nodes. This happened for LaFeAs(O,F) (critical field [2], μSR [3] NMR [4], specific heat [5], point-contact spectroscopy [6]), SmFeAs(O,F) (tunnel and point-contact spectroscopy [7],[8]), (Sr,K)Fe$_2$As$_2$ (STM [9]), LaFePO (specific heat [10]) and some others.

Although the *d*-wave symmetry is theoretically compatible with the shape of the Fermi surface (FS) and with the multiple spin-fluctuation modes arising from the nesting across its disconnected sheets [11], the occurrence of line nodes crossing the Fermi surface has been later contradicted by direct ARPES measurements in single crystals of compounds belonging to the 122 and the 1111 family, i.e. (Ba,K)Fe$_2$As$_2$ [12],[13] and NdFeAs(O,F) [14] .

In the compounds of the 122 family, ARPES shows that a nodeless gap opens up below T$_c$ over the whole Fermi surface, but its amplitude depends on the sheet of the FS. The gap amplitudes are actually clustered around two well-distinct values, $\Delta_1$ = 5.8 meV and $\Delta_2$ = 12 meV. The presence of multiple isotropic gaps in these compounds is also particularly clear in a variety of experiments such as infrared spectroscopy [15], specific heat [16] and lower critical field [17] measurements, point-contact spectroscopy [18] and so on. Whether this evidence is compatible with the proposed s± scenario [19] is however still to be clarified, since no unambiguous information has been obtained up to now about the phase of the order parameter.

The case of the 1111 family is less clear. Here ARPES measurements have been performed up to now on one single FS sheet [14] (where a nodeless gap was observed), but multiband superconductivity is supported by critical field measurements in LaFeAs(O,F) [20] as well as by high-field vortex torque magnetometry [21] and penetration depth measurements [22] in SmFeAs(O,F) single crystals. The presence of *at least* one nodeless gap (whose amplitude is approximately BCS) was also shown by point-contact spectroscopy measurements in SmFeAs(O,F) [23] while the same technique applied to NdFeAs(O,F) [24] and oxygen-deficient NdFeAsO [25] gave complex results very difficult to interpret. Recent STM measurements on Sm-1111 single crystals [26] showed the presence of large-energy features in addition to the superconducting gap. They were interpreted as being the hallmark of the coupling of quasiparticles to a collective spin excitation, but may also indicate the presence of a second gap spatially coexisting with the smaller one. The c-axis STM spectra look indeed very similar to the unnormalized one reported by Wang et al [8] (fig.4e) apart from the zero-bias conductance peak that turns out to be an artifact of the point-contact technique. Two gaps ($\Delta_1$ ~ 9 meV and $\Delta_2$ ~ 18 meV) were also observed by STM in NdFeAs(O,F), but in different spatial locations [27].

In this paper we report on the results of point-contact spectroscopy measurements in polycrystalline samples of two compounds of the 1111 family, LaFeAsO$_{1-x}$F$_x$ (La-1111) and SmFeAsO$_{1-x}$F$_x$ (Sm-1111). The spectra contain clear and reproducible evidence of two distinct sets of features, such as low-energy conductance peaks and higher energy peaks or shoulders, that make the spectra look very similar to those measured in MgB$_2$. Here we will interpret both these sets of features as being the hallmark of two different superconducting order parameters, possibly opened on different sheets of the Fermi surface as it happens in the 122 compounds. It should be borne in mind, however, that a different interpretation (where the smaller energy scale is a superconducting gap while the second is something else) is possible anyway, as shown in [28]. In any case both these energy scales are related to superconductivity and do not exist in the normal state. Possibly because of the pressure-less point-contact technique we used, our spectra do not show zero-bias conductance peaks so that the *d*-wave symmetry of the gap is definitely ruled out. We will show that, while in Sm-1111 both the gaps close at the bulk T$_c$ of the samples and approximately follow a BCS-like temperature dependence, in La-1111 the larger gap seems to close below T$_c$ and the small one presents a high-temperature "tail" which is absolutely non-conventional. In both cases, the small gap $\Delta_1$ is smaller than the *s*-wave weak coupling value ($2\Delta_1$/ k$_B$T$_c$ = 2.2 – 3.2) while the second is by far larger than that, with a gap ratio $2\Delta_2$/ k$_B$T$_c$ = 6.5 – 9.

## 2. Sample preparation and characterization

The polycrystalline samples used in the present work were grown by solid-state reaction using two different routes. LaFeAsO$_{1-x}$F$_x$ samples with nominal F content $x$ = 0.1 were obtained starting from a mixture of the four precursor materials (LaAs, Fe$_2$O$_3$, Fe, and LaF$_3$) that were ground and cold-pressed into pellets. Then, they were placed into a Ta crucible, sealed in a quartz tube under argon atmosphere, and annealed at ambient pressure and 1150°C for 50 h. SmFeAsO$_{1-x}$F$_x$ samples with $x$=0.2 were instead grown starting from SmAs, FeAs, SmF$_3$, Fe$_2$O$_3$ and Fe by means of a high-pressure synthesis. The pulverized starting materials were sealed in a BN crucible, brought to a pressure of 30 kbar at room temperature, heated within 1 h up to 1350-1450 °C, kept for 4.5 h at this temperature and finally quenched back to room temperature.

The resulting La-1111 samples show a disordered matrix where large crystallites of 5 – 20 μm are immersed, while the Sm-1111 samples are more compact and contain shiny crystallites whose size is of the order of 30 μm. In both cases the size of the crystallites was determined by SEM images and micro energy-dispersive X-ray spectroscopy (EDX) was used to measure the local F content. It results quite homogeneous in Sm-1111 samples, while in La-1111 ones it is uniform inside each crystallite but shows variations up to $\Delta x$ = 0.02 from one crystallite to another. Figure 1 shows the resistivity of La-1111 samples (a) and of Sm-1111 ones (b) as a function of temperature. The onset critical temperature T$_c^{on}$ (defined at 90% of the resistive transition) is 53 K for Sm-1111 and 27 K for La-1111 while the transition widths (10%-90% of the resistive transition) are $\Delta T_c$ ~ 2 K and 4 K, respectively. The resistivity of La-1111 samples clearly exhibits a deviation from the low-temperature normal-state behaviour already at a temperature of the order of 31 K (indicated by a vertical dashed line in Fig. 1(a)). This fact appears consistent with the observed variations of the local F content measured by EDX in these samples and with the variations of the local critical temperature of the point-contact Andreev-reflection curves (T$_c^A$). As a matter of fact, all the T$_c^A$ values observed in point contacts on our LaFeAsO$_{1-x}$F$_x$ samples lie in the region between the two vertical dashed lines of Fig. 1(a) i.e. between 27 K and 31 K. Instead, all the T$_c^A$ values of the SmFeAsO$_{0.8}$F$_{0.2}$ contacts are within 1 K of the bulk T$_c$.

## 3. Point-contact Andreev-reflection measurements

In order to obtain our point contacts, we first break or splinter our samples at room temperature in order to expose a fresh surface. We then immediately create the contact by placing a small drop (diameter ∅ < 50 μm) of Ag conductive paint (containing Ag grains 2-10 μm in size) on the new surface. This "soft" point-contact technique, alternative to the standard one (where a sharp metallic tip is pressed against the sample surface), ensures a very good mechanical and thermal stability of the contacts and eliminates the thermal drifts of the tip. This prevents the contact change at the increase of temperature and allows easily reaching high temperatures (200 K and over) with stable contacts [29]. Moreover, this pressure-less technique eliminates possible pressure-induced lattice distortions in the region of the contact. The last two points appear particularly important in pnictides where the possibility to measure stable PCAR curves at temperatures higher than the Neél temperature of the parent compound is very useful and where the emergence of zero-bias anomalies can be ascribed to a systematic effect of the tip pressure [23], [25]. In a way totally similar to what happens with the standard PCAR technique, the apparent size of the macroscopic contact (here the diameter of the Ag-paint spot, there the diameter of the deformed top of the tip), however, does not correspond to the actual size of the nanoscopic point contacts created on the sample surface. As a matter of fact, parallel microjunctions are very likely to form between the sample surface and the Ag particles in the paint, so that the measured I-V characteristics and conductance curves should be regarded as an average over a certain region in direct space. The inhomogeneous distribution of F in the different crystallites of the La-1111 sample together with their relatively small size – when compared to the macroscopic size of the point contact – suggest some care in interpreting the experimental data in this material. In fact, in this case, the occurrence of multiple parallel ballistic point contacts on crystallites with different $T_c$ and gap(s) cannot be *a priori* excluded. We have simulated this situation by averaging the theoretical Andreev-reflection (AR) curves calculated by the generalized Blonder-Tinkham-Klapwijk model using the parameters typical of experimental data and the maximum and minimum $\Delta$ and $\Gamma$ values observed in our true contacts. The resulting averaged curve can be perfectly fitted using $\Delta$ and $\Gamma$ values that are the average of the starting ones. This fact suggests that even the maximum doping inhomogeneity of our La-1111 samples does not have a dramatic effect on the shape of the conductance curves simply resulting, in the worst case, in an average of the superconducting gap value(s) over a few crystallites. The main consequence is thus an increase of the uncertainty in the determination of the gap(s) (particularly the large one), provided that the gap amplitudes are reported as a function of the local critical temperature in the region of the contact, $T_c^A$ and not as a function of the bulk $T_c$, as it has been already demonstrated in doped $MgB_2$ [30], [31]. Additional details can be found in literature [28], [32].

Usually, the normal-state resistance of the freshly-made contacts ($R_N$) is lower than 100 $\Omega$ and they are in the Andreev-reflection regime. If the initial resistance of the as-made contact is too high, it can be tuned by means of short voltage or current pulses, as described for example in Ref. [29] and as well known since the Seventies [33]. Nevertheless, during these pulses the contact region can be heated for a short time well above the bath (or even room) temperature possibly producing modifications of the sample surface. For this reason in the present paper we consider only conductance curves coming from as-made contacts. In the past seven years the "soft" point-contact technique just described proved to be very effective and reliable in the determination of the gaps and their temperature and magnetic-field dependence in two-band and anisotropic superconductors [29], [31], [34].

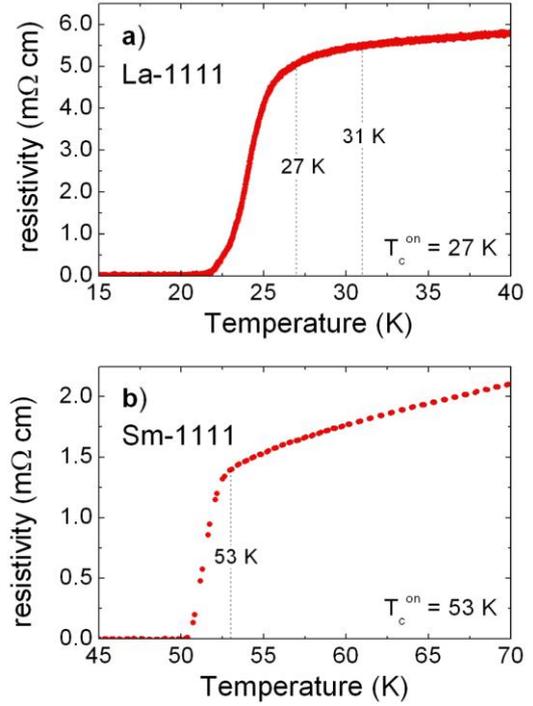

**Figure 1.** (a) Resistivity of the La-1111 sample as a function of temperature, in the region of the superconducting transition. Vertical dashed lines indicate the initial deviation from the low-temperature linear behaviour of the resistivity, occurring at 31 K, and the critical temperature $T_c^{on}$=27 K, defined at 90% of the transition. (b) The resistivity of the Sm-1111 sample. Here $T_c^{on}$ (defined as in panel (a) and indicated by the vertical dashed line) is 53 K.

We have seen that $R_N$ has to be small for the contact to be in Andreev-reflection regime but, at the same time, it cannot be too small to fulfil the condition for ballistic transport through the junction ($a \ll \ell$, being $a$ the contact radius and $\ell$ the electronic mean free path), so that charge carriers are not scattered in the contact area. As a matter of fact, in a perfectly ballistic contact $a$ depends on $R_N$ according to the Sharvin formula, so that the greater is $R_N$, the smaller is $a$ and, therefore, the previous condition on $a$ turns into a condition on $R_N$, which must exceed a minimum value related to $\ell$. Unfortunately an estimation of $\ell$ in these materials is very difficult and uncertain due to the small density of charge carriers [35] and only the reproducible absence in the dI/dV curves of sharp dips or other signs of heating effects can *a posteriori* guarantee the presence of ballistic conduction in the contact.

The differential conductance curves (dI/dV vs. V, where V is negative when electrons are injected into the superconductor) were obtained by numerical differentiation of the measured I-V characteristics. As usual we selected only the curves with clear Andreev-reflection features (usually those with $R_N$ < 100 $\Omega$) and measured the full temperature dependence of these conductances (in these materials up to ~ 200 K) so as to evaluate the temperature (previously called $T_c^A$) at which the Andreev-reflection features disappear and the normal-state conductance is recovered, and to study the behaviour of dI/dV at $T > T_c^A$.

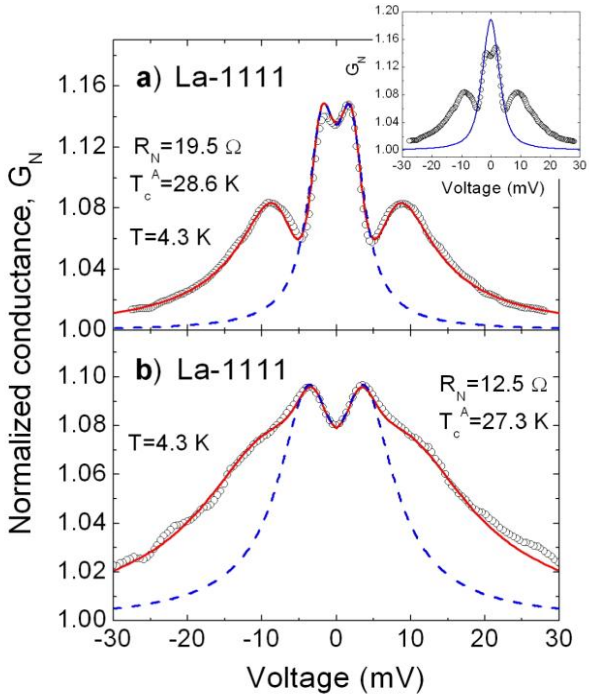
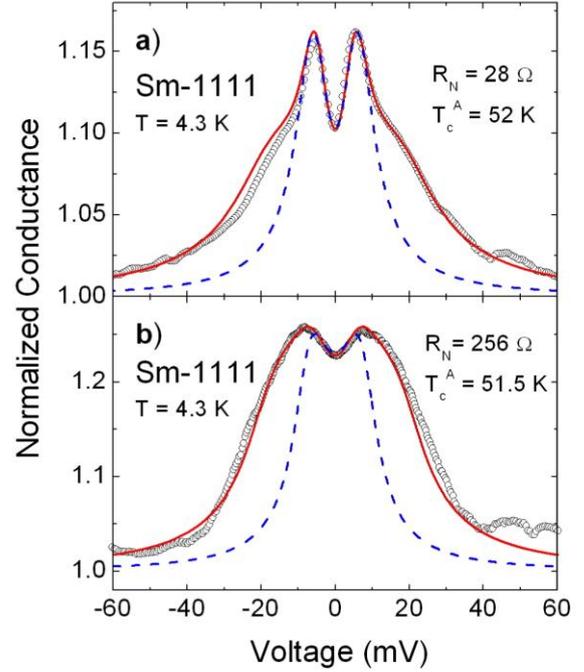

**Figure 2.** (a) Circles: normalized conductance curve measured at 4.3 K in La-1111 in a point contact with normal-state resistance $R_N$=19.5 $\Omega$ and critical temperature $T_c^A$=28.6 K. Lines represent the best-fitting s-wave 3D generalized BTK curves, in the single-band (dashed line) and two-band (solid line) case. Inset: the same experimental data compared to the 3D generalized BTK *d*-wave fit (line) averaged over all possible values of $\beta \in [0, \pi/4]$ ($\beta$ being the angle between the normal to the interface and the antinodal direction). (b) same as in (a), but for a 12.5 $\Omega$ contact with $T_c^A$=27.3 K.

**Figure 3.** (a) Normalized conductance curve (measured at 4.3 K) of a Ag/Sm-1111 point contact with $R_N$=28 $\Omega$ and $T_c^A$=52 K. Lines represent the best-fitting s-wave 3D generalized BTK curves, in the single-band (dashed line) and two-band (solid line) case. (b) Same as in (a) but for a different point contact with $R_N$=256 $\Omega$ and $T_c^A$=51.5 K.

The main goal of this paper is to convince the reader of the *clear and reproducible presence* in our point-contact Andreev-reflection data in both LaFeAsO$_{1-x}$F$_x$ and SmFeAsO$_{0.8}$F$_{0.2}$ of spectroscopic features that can be explained and fitted *only* by considering the presence of *two nodeless gaps* in the quasiparticle excitation spectra of these materials. As a consequence, we first concentrate on this goal by directly showing in Fig. 2 and 3 some examples of the normalized conductance curves we measured in both La-1111 and Sm-1111 at low temperature (open symbols). Of course this way to present the data pushes the very important problem of the normalization of the dI/dV curves somewhat into the background.

Given that the full details on the normalization of our curves (including a comparison between two different procedures) are present in literature [28], [32], here we can simply say that, due to the very high upper critical fields of these superconductors, we cannot have experimental access to the low-temperature normal-state conductance of the samples. As a consequence, the only two ways to normalize the dI/dV curves rest on the use for the normal-state background of the measured conductance at $T = T_c^A$ (vertically translated if necessary) [28], [32] or of a proper smooth curve that connects the high-bias tails of the conductance curves with a suitable point at zero bias [28]. In both materials the division by a normal background either constant or linearly dependent on energy is not possible due to the peculiar shape of the conductance curves at $T = T_c^A$ which present a marked depression (La-1111) or a hump (Sm-1111) at zero bias, as it will be shown and discussed in the final part of this article. Moreover, we have proved that the gaps and their temperature dependence extracted from the conductance curves are practically independent of the two different normalization procedures [28], allowing us to concentrate on the physical results present in the normalized dI/dV curves.

Figure 2 shows the normalized conductances at 4.3 K (open symbols) of two different point contacts on La-1111 having $R_N$ = 19.5 $\Omega$ and $T_c^A$ = 28.6 K (panel a) and $R_N$ = 12.5 $\Omega$ and $T_c^A$ = 27.3 K (panel b). Figure 3 shows the normalized dI/dV curves at the same temperature (open symbols) for two different point contacts in Sm-1111 having $R_N$ = 28 $\Omega$ and $T_c^A$ = 52 K (panel a) and $R_N$ = 256 $\Omega$ and $T_c^A$ = 51.5 K (panel b). Simply by looking at these curves several typical features can be evidenced.

First, no conductance peaks at zero bias (ZBCP) are observed. This finding is common to all the PCAR curves we measured in La-1111 and in Sm-1111 and clearly suggests that the ZBCPs observed in some point-contact spectroscopy (PCS) measurements in iron oxypnictides [6], [8], [23], [24], [25] are probably an extrinsic feature of PCS spectra, related to the pressure applied by the tip in the conventional technique [25] and mainly appearing in junctions with a small normal-state resistance [23], [36].

Together with the polycrystalline nature of our samples, these results prove the *absence of nodes* in the superconducting gap of these two Fe-based pnictides. As a matter of fact, it can be easily shown by numerical simulations of AR curves within the generalized Blonder-Tinkham-Klapwijk (BTK) model [37] that a *d*-wave symmetry of the superconducting order parameter would lead to ZBCP whenever the angle β between the normal to the NS interface and the anti-nodal direction is larger than π/16. It means that for the parameters of our best PCAR curves, i.e. a ratio of the gap over the broadening parameter $\Delta(0)/\Gamma(0) \sim 2$ and a barrier parameter $Z \sim 0.3-0.4$ (see the following for details), the normalized conductance in *d*-wave symmetry shows a two-peak structure similar to the experimental one only for β < π/16. It is impossible to believe that *all* the random contacts in our polycrystalline samples have "magically" selected current-injection directions with β < π/16 with respect to the randomly oriented crystallites. To further convince the reader of this fact, in the inset of Fig. 2 (a) we show a tentative fit of the central part of the normalized conductance by means of the 3D BTK model in *d*-wave symmetry [37], averaging over all β ∈ [0, π/4] (solid line). As expected, the resulting AR conductance shows a peak at zero bias that is completely different from the actual and reproducible experimental findings.

The experimental AR curves of *all* the contacts in La-1111 and Sm-1111 show two clear peaks at low bias, plus additional structures (peaks or shoulders) at a higher voltage. In La-1111 these low-bias peaks are at about 1.7 – 3 meV (see Fig. 2 (a) and (b)), while in Sm-1111 they are at about 5.5 – 8 meV (see Fig. 3 (a) and (b)). In Fig. 2 (a) the additional high-bias structures assume the form of true conductance peaks at ± 8.9 meV, while in the other contacts of Fig. 2 and 3 they simply show up as shoulders at about ±10 meV (in La-1111, Fig. 2 (b)) or at about ±14 meV (in Sm-1111, Fig. 3 (a) and (b)). All these dI/dV curves appear quite similar to the PCAR conductances measured in the past seven years by us and by many other authors in $MgB_2$ and related compounds [29], [38], [39], where the aforementioned features have been attributed to the presence of a two-gap superconductivity in the material. Several papers have indeed recently suggested the presence of a two-gap superconductivity in Fe-based pnictides on the basis of critical-field measurements in La-1111 [20], point-contact spectroscopy and torque magnetometry in Sm-1111 [8], [40], as well as angle-resolved photoemission spectroscopy (ARPES) [12], [41], point-contact spectroscopy [18] and penetration depth measurements [42] in $Ba_{0.6}K_{0.4}Fe_2As_2$. The Fermi surfaces of all these materials share quite common features exhibiting separate hole-like and electron-like sheets that, if the superconducting properties are determined by the interband coupling, could lead to multiple-gap superconductivity [43].

Let us forget for a moment these evidences and, as a first approximation, try to fit our PCAR curves with a single-gap BTK model generalized to take into account broadening effects of various nature [29], [34], [44] and the angular distribution of the injected current at the interface [37]. The free parameters of this model are the energy gap Δ, the parameter Z (that depends on the height of the potential barrier at the NS interface) and the broadening parameter Γ. The results of this single-gap fit are shown as dashed curves in Fig. 2 and 3. It is quite clear that both in La-1111 and in Sm-1111 this generalized single-gap BTK model is only able to properly fit the low-bias region of the spectra where the two main peaks are present. The parameters obtained from the fit of the two PCAR curves on La-1111 are: Δ = 2.15 meV, Γ = 1.39 meV, Z = 0.33 (Fig. 2 (a)) and Δ = 3.90 meV, Γ = 3.48 meV, Z = 0.36 (Fig. 2 (b)), while those obtained in Sm-1111 from the curves of Fig. 3 (a) and (b) are Δ = 6.45 meV, Γ = 3.80 meV, Z = 0.38 and Δ = 8.0 meV, Γ = 3.7 meV, Z = 0.27, respectively. However, even at a first glance, it is evident that the generalized single-gap BTK model is totally unsuitable for properly fitting the additional structures at higher bias. These structures are not the sharp spikes or dips that sometimes appear at casual voltage in the conductance curves of non-ideal contacts. They appear in all the point contacts on the same superconductor at about the same energy and show a common behaviour shifting in energy and decreasing in amplitude at the increase of temperature. All the previous considerations as well as our experience of point-contact spectroscopy in pure and doped $MgB_2$ tell us that only the presence of a second nodeless order parameter can reasonably explain the features observed at high bias. As a consequence, we fitted the low-temperature normalized dI/dV curves with the two-gap version of the generalized BTK model. According to this version, the normalized point-contact conductance $G_N$ is the weighted sum of two generalized single-gap BTK contributions, i.e. $G_N(E) = w_1 G_1^{BTK}(E) + (1-w_1) G_2^{BTK}(E)$. Now the parameters of this generalized two-gap BTK model are the gap magnitudes $\Delta_1$ and $\Delta_2$, the potential barrier parameters $Z_1$ and $Z_2$, the broadening parameters $\Gamma_1$ and $\Gamma_2$, plus the weight $w_1$. Of course, by using this model we implicitly assume that both the order parameters evidenced by our PCAR curves have a superconducting origin and that both the bands contribute independently to the total conductance, completely neglecting, as a first approximation, the band mixing due to the interband scattering at the surface. The results of this two-gap fit are shown as solid lines in Fig. 2 and 3. It is evident that now the fit can almost perfectly follow the whole dI/dV curves when either additional peaks or shoulders are present at high bias.

| Material | $T_c^A$ (K) | $\Delta_1$ (meV) | $\Gamma_1$ (meV) | $Z_1$ | $\Delta_2$ (meV) | $\Gamma_2$ (meV) | $Z_2$ | $w_1$ |
|---|---|---|---|---|---|---|---|---|
| $LaFeAsO_{1-x}F_x$ (Fig. 2 a) | 28.6 | 2.75 | 0.95 | 0.21 | 7.9 | 3.9 | 0.93 | 0.6 |
| $LaFeAsO_{1-x}F_x$ (Fig. 2 b) | 27.3 | 3.8 | 3.05 | 0.28 | 10.2 | 7.0 | 0.49 | 0.6 |
| $SmFeAsO_{0.8}F_{0.2}$ (Fig. 3 a) | 52 | 5.7 | 3.23 | 0.42 | 19.0 | 9.05 | 0.3 | 0.65 |
| $SmFeAsO_{0.8}F_{0.2}$ (Fig. 3 b) | 51.5 | 6.6 | 5.27 | 0.45 | 20.0 | 6.7 | 0.16 | 0.48 |

**Table 1**
Critical temperatures of the point contacts and parameters of the generalized two-band BTK model obtained from the fits of the curves of Fig. 2 and 3.

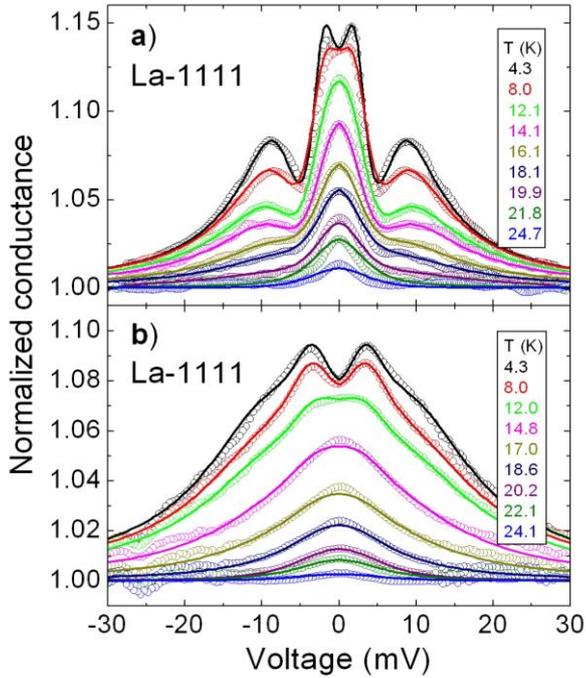

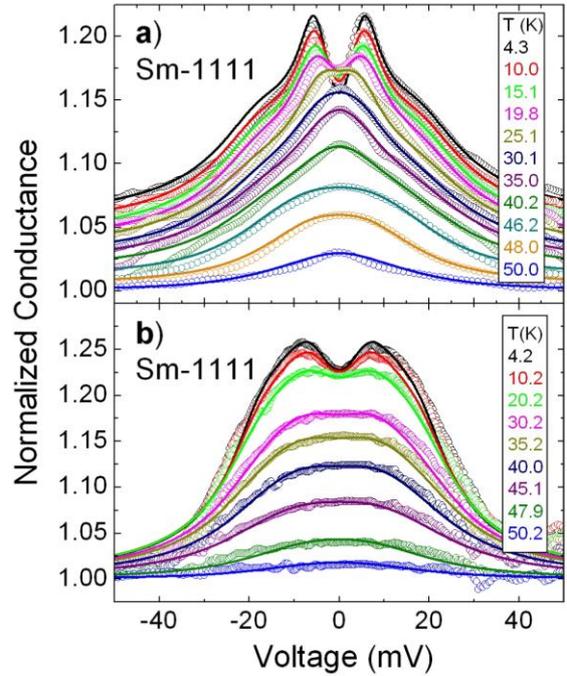

**Figure 4.** The conductance curves of the same Ag/La-1111 contacts as in Fig.2a and 2b, respectively, measured at different temperatures up to ~ $T_c^A$ (symbols). The top curve in each panel is measured at 4.3 K, the others (from top to bottom) at the temperatures indicated in the legend. Lines represent the two-band, generalized s-wave BTK fit of the experimental data.

**Figure 5.** The conductance curves of the same Ag/Sm-1111 contacts as in Fig.3a and 3b, respectively, measured at different temperatures up to ~ $T_c^A$ (symbols). The top curve in each panel is measured at 4.3 K, the others (from top to bottom) at the temperatures indicated in the legend. Lines represent the two-band, generalized s-wave BTK fit of the experimental data.

The values of the parameters of the generalized two-band BTK model obtained from the fits of the curves of Fig. 2 and 3 are summarized in Table 1.

In the best contacts the ratios $\Delta_1/\Gamma_1$ and $\Delta_2/\Gamma_2$ are of the order of (or greater than) 2 and the weight of the conductance related to the small gap is always $w_1 \sim 0.5 - 0.65$. However, some contacts present smaller ratios and, generally speaking, the $\Gamma$ values are rather large in comparison with the intrinsic lifetime broadening expected in these materials. This situation is common to all the PCAR experiments we made in the past eight years on different materials by using the "soft" technique and did not prevent us from obtaining gap values very reliable and in very good agreement with theoretical calculations [29], [31], [34]. After some experimental investigations on the chemistry of the Ag grains, we concluded that the most probable cause of the small amplitude of our conductance curves and, as a consequence, of the large $\Gamma$ values, is the presence of a thin layer on the surface of the Ag grains that gives rise to quasiparticle inelastic scattering, similarly to what shown in Ref. [45].

Having proved that the generalized two-gap BTK model is very effective in fitting the PCAR curves both in La-1111 and in Sm-1111, we are now ready to apply it to the whole temperature dependency of the conductances shown in Figs. 2 and 3. The results are presented in Figs. 4 and 5 where experimental data are shown as open circles and BTK fitting curves as solid lines. In all cases the two-gap fit is very good at any temperature but, in many contacts (particularly in Sm-1111), a residual asymmetry of the normalized curves with respect to the sign of the bias is present. This asymmetry could be an intrinsic feature of the superconducting state in these materials, but, more likely, it is due to the shape of the (inaccessible) low-temperature normal-state conductance which may be more asymmetric than that measured at $T_c^A$. For this reason, when necessary, we fitted both sides of the normalized curves in order to determine the spread of $\Delta$ values arising from this asymmetry. For example, the BTK curves shown in Fig. 5 refer to the fit of experimental data for positive (a) and negative (b) bias. It is important to remember that not all the seven fitting parameters of the model are totally free. As a matter of fact the two barrier parameters $Z_1$ and $Z_2$ as well as the weight $w_1$ should remain (almost) constant with increasing temperature. The broadening parameters $\Gamma_1$ and $\Gamma_2$ should also remain almost constant, or, at most, increase with temperature. Of course, these conditions automatically restrict the variability of these parameters.

The temperature dependencies of the energy gaps of La-1111 obtained from the fits of the dI/dV curves of Fig. 4 are reported in Fig. 6 a (open and solid symbols). The corresponding broadening parameters remain almost constant ($\Gamma_1$) or show a regular increase with temperature ($\Gamma_2$) of the order of 22 – 50 %, while the barrier parameters $Z_1$ and $Z_2$ and the weight $w_1$ are temperature independent thus being fixed at the values indicated in Table 1.

Two striking characteristics are evident in these curves. First, the gap values $\Delta_1$ and $\Delta_2$ (particularly the latter) are sensibly different in the two contacts on the same sample. This apparent inconsistency is actually related to the observed dependency of both the gaps on the critical temperature of the junction, $T_c^A$, i.e. on the *local* $T_c$ of the material at the point where the contact occurs. We have already shown in the past that in rather inhomogeneous samples (e.g. highly-doped $MgB_2$) $T_c^A$ can vary up to some degrees from contact to contact on the same sample, due to local variations in the doping content. In this situation the gaps obtained by PCAR spectroscopy are related to $T_c^A$ rather than to the bulk $T_c$ of the sample [46]. We already pointed out that the local F content in the different crystallites of our La-1111 samples shows variations up to 20% that are compatible with the observed variations of $T_c^A$ (between 27.3 and 31 K) as well as with the shape of the resistivity shown in Fig. 1 (a). In these contacts we reproducibly observed that both the gaps depend on $T_c^A$. $\Delta_1$ increases with $T_c^A$, always being smaller than the $s$-wave BCS value, while $\Delta_2$ is very large and rapidly decreases at the increase of $T_c^A$ [28]. A detailed discussion on the dependence of the gaps from the local critical temperature of the contacts in La-1111 samples is beyond the scope of the present paper but can be found in Ref. [28].

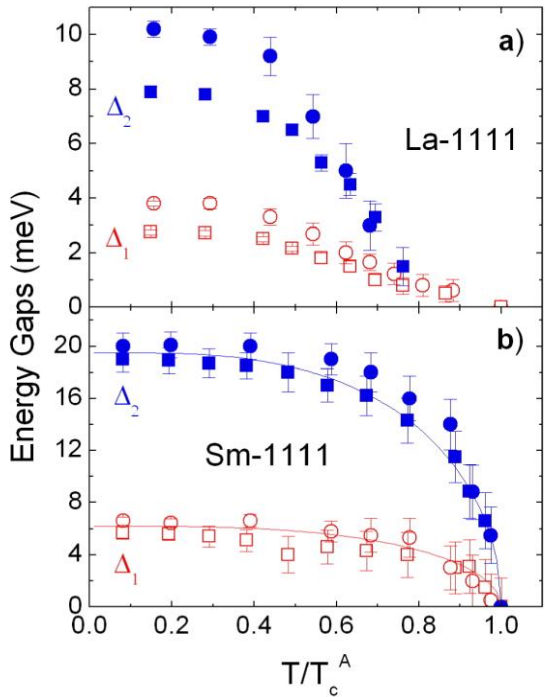

**Figure 6.** (a) The gaps $\Delta_1$ (open symbols) and $\Delta_2$ (full symbols) as obtained from the fit of the conductance curves reported in Fig.4a (squares) and 4b (circles). The data are shown as a function of the normalized critical temperature $T/T_c^A$ to allow a direct comparison between the trends obtained in different contacts. (b) The gaps $\Delta_1$ and $\Delta_2$ as a function of the normalized temperature $T/T_c^A$ as obtained from the fit of the conductance curves in Fig.5a (squares) and 5b (circles). Lines are BCS-like curves. The error bars take into account the uncertainty due to the fitting procedure.

The second characteristic of the curves shown in Fig. 6 (a) is a definitely non BCS temperature dependency of the gaps that smoothly and regularly decrease on increasing T, but show large deviations from the BCS-like behaviour at T > 0.5 $T_c^A$. Both the small and the large gap decrease at the increase of temperature faster than expected in a BCS framework for $T_c^A \sim 27$ K, and are clearly suppressed above T ~ 0.8 $T_c^A$. Above this temperature, the smaller gap, $\Delta_1$, exhibits a "tail" up to the critical temperature of the contact, while the larger one, $\Delta_2$, either disappears or becomes so small that it is impossible to detect it within our experimental resolution. This does not exclude that also $\Delta_2$ shows a tail up to $T_c^A$; but, given the large value of $\Gamma_2$, it would be impossible to clearly discern signs of its presence in the conductance curves. The possibility that the tail is ascribed to the large gap $\Delta_2$ while $\Delta_1$ disappears at T ~ 0.8 $T_c^A$ can be instead excluded. As a matter of fact, fitting the conductance curves in the proximity of $T_c^A$ with $\Delta_2$, $\Gamma_2$ and $Z_2$ as the only parameters would require a sudden reduction of $\Gamma_2$ and $Z_2$ which appears to be non physical. If the tail is instead due to $\Delta_1$, the fit is possible without any discontinuity in the fitting parameters $\Gamma_1$ and $Z_1$. At the present moment the reason of this anomalous behaviour is not clear but, from the experimental point of view, we have evidence it is reproducibly present in all the contacts we measured in La-1111, it does not depend on the choice of the normalization procedure and it persists even in the case we fit only the central part of the conductance curves by a single-gap BTK model (as we did in Fig. 2 and 3, dashed lines) [28]. Further measurements in highly homogeneous samples may help finally clarifying this point.

Figure 6 (b) shows the temperature dependencies of the energy gaps of Sm-1111 (open and solid symbols) obtained from the fits of the PCAR conductance curves of Fig. 5 (a) (positive bias) and 5 (b) (negative bias). Also in this case, $\Gamma_1$ remains almost constant (within ± 15 − 20 %) at the increase of temperature, while $\Gamma_2$ remains almost constant or increases regularly. A small change in $Z_1$ and $Z_2$ (of the order of 20%) has to be allowed in order to fit properly the whole temperature dependencies while $w_1$ is kept at the low-temperature value shown in Table 1.

These temperature dependencies are clearly much more regular than those reported in Fig. 6 (a) showing, in both cases, two energy gaps that close at the same temperature $T_c^A$ (52 K and 51.5 K, in the two contacts) very close to the bulk $T_c$ of the sample and follow very well a BCS-like behaviour. In this case the values of $\Delta_1$ (or $\Delta_2$) obtained in different contacts are close to each other being separated by ~ 1 meV at low T, but, particularly for $\Delta_2$, this is not true in all the contacts we studied. In fact various other measurements in different contacts have confirmed the nice BCS-like trend of both the gaps but have also shown that, while the $\Delta_1(0)$ values are very reproducible, giving an average value at low temperature $\Delta_1(0) = 6.15 \pm 0.45$ meV, the $\Delta_2(0)$ values are more widely spread between the different data sets (from 15 meV to ~ 21 meV) leading to an average value $\Delta_2(0) = 18 \pm 3$ meV [32]. This unusual spread, never observed in previous PCAR measurements by using the same "soft" technique, is mainly due to the residual asymmetry of the normalized curves with respect to the bias voltage (shown, for example, in Figs. 3 (a) and 5 (a)). We already pointed out that this asymmetry is likely to be ascribed to the anomalous shape of the normal-state conductance that here presents a large hump at zero bias (see Fig. 7 (b)), even if the possibility that it is an intrinsic feature of the superconducting state in this material cannot be ruled out.

For a moment let us forget the quite different temperature dependence of the gaps we observed in La-1111 and Sm-1111 samples and concentrate on their low-temperature values $\Delta_1(0)$ and $\Delta_2(0)$. From the data of Table

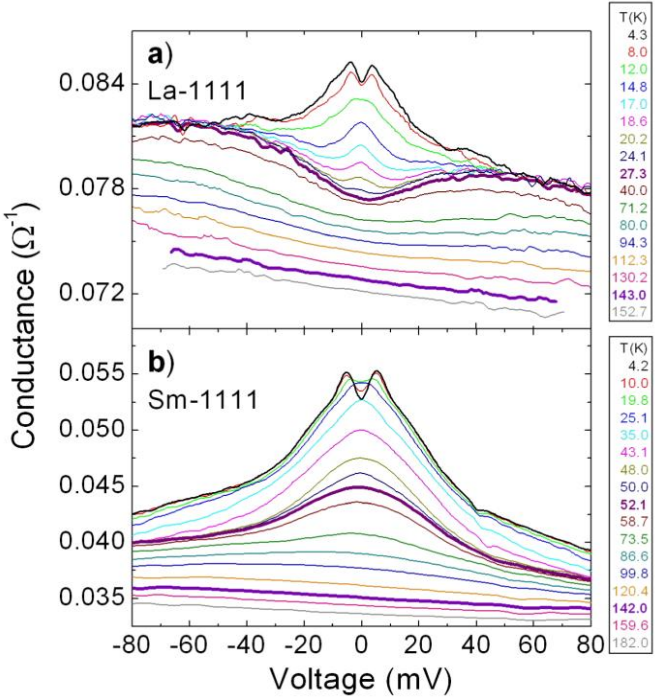

**Figure 7.** (a) The raw (unnormalized) conductance curves of the same Ag/La-1111 contact as in Fig. 2b and 4b, measured from 4.3 K (top curve) up to 153 K (bottom curve). The curves clearly show the persistence of a depression in the conductance in the normal state, which is progressively filled on increasing temperature and disappears around 140 K. (b) Raw conductance curves of the same Ag/Sm-1111 contact as in Fig. 3a and 5a, measured from 4.2 K up to 182 K. In this case, the normal state shows a broad zero-bias hump that is progressively washed out on increasing the temperature and disappears at about 140 K.

1 we obtain that the ratio $2\Delta_1(0)/k_BT_c^A$ is in the range 2.2 – 3.2 for La-1111 and 2.5 – 3.0 for Sm-1111, while $2\Delta_2(0)/k_BT_c^A$ values span between 6.4 and 8.7 in La-1111 and between 8.5 and 9 in Sm-1111. If we consider all the PCAR curves measured in Sm-1111 we simply observe an increase of the spread of $2\Delta_2(0)/k_BT_c^A$ that spans between 7 and 9. The ratio $\Delta_2(0)/\Delta_1(0)$ is always clse to 3 being equal to 2.7 – 2.9 in La-1111 and 3.0 – 3.4 in Sm-1111. In our opinion these numbers are quite interesting since they demonstrate, independently of the structural and physical differences between these two compounds, that they apparently share some common properties: A small nodeless gap with a ratio $2\Delta_1(0)/k_BT_c^A$ lower than the BCS value and a large, also nodeless, gap with $2\Delta_2(0)/k_BT_c^A$ much (up to more than 2.5 times) higher than the s-wave BCS one.

Finally let us go back to the raw PCAR data just for showing the evolution of the conductance curves up to T ≈ 200 K in both the La-1111 samples and in the Sm-1111 ones. This is done in Fig. 7 (a) and (b), respectively. The normal-state conductance at $T_c$ in La-1111 (upper thick line in Fig. 7 (a)) is asymmetric and shows a depletion of the density of states around the Fermi level with two broad humps at energies of the order of 50 – 60 meV and a broad minimum at zero bias. These structures, observed in all the contacts on La-1111 samples, persist below $T_c$ coexisting with superconductivity [28] and progressively smooth out on increasing the temperature, finally disappearing at about 140 K and leaving a flat but still asymmetric conductance (bottom thick line in Fig. 7 (a)). This pseudogap-like shape of the normal-state conductance is very similar to that observed by PCS in $URu_2Si_2$, a material with long-range spin-density-wave (SDW) order [47]. This static long-range SDW order is certainly not present in superconducting La-1111 [3]. On the other hand, i) the disappearance of the pseudogap-like features at a temperature very close to the Néel temperature of the antiferromagnetic (AF) SDW state in the parent compound [48], ii) the sensitivity of PCAR spectroscopy to the electron dynamics on a short time scale, and iii) recent theoretical considerations concerning the opening of a pseudogap in 2D systems in the presence of local AF fluctuations, all suggest the existence of spin fluctuations coexisting with superconductivity in the doped compound at low temperature. Further details can be found in literature [28].

As far as Sm-1111 is concerned, Fig. 7 (b) shows the evolution of the raw conductance of the contact already presented in Figs. 3 (a) and 5 (a) at the increase of temperature up to about 180 K. Unlike in La-1111, here the normal-state conductance measured at $T_c$ (upper thick line in Fig. 7 (b)) shows a hump at zero bias that gradually decreases at the increase of temperature until it completely disappears again at T ~ 140 K (bottom thick line in Fig. 7 (b)).

We have no clear explanation for this anomalous normal-state conductance but we would like to point out that similar features of the normal-state spectrum have been recently observed in $Ba_{0.6}K_{0.4}Fe_2As_2$ by point-contact spectroscopy [18] and ARPES [12] measurements. Again, the proximity of the disappearance of this hump to the Néel temperature of the parent compound suggests its possible magnetic origin, but further theoretical and experimental work has to be done in order to understand the reasons for such an anomalous behaviour.

## 4. Conclusions

In this paper we have reported the results of point-contact Andreev-reflection spectroscopy measurements in polycrystalline samples of two Fe-based superconductors of the 1111 family, i.e. $LaFeAsO_{1-x}F_x$ (with nominal x = 0.1) and $SmFeAsO_{1-x}F_x$ (with x = 0.2). We have shown that, in both cases, the PCAR spectra show clear peaks related to a superconducting gap, plus additional features at higher bias, in the form of peaks or shoulders. The shape of the spectra is in most cases very similar to that of analogous spectra measured in the two-band system $MgB_2$. They instead never show zero-bias conductance peaks. Taking into account the polycrystalline nature of the samples, the non-perfectly directional current injection in PCAR experiments and the irregular normal metal/superconductor interface, the absence of ZBCP definitely rules out the possibility of a d-wave symmetry of the superconducting order parameter – or, more generally, the possibility of line nodes crossing the Fermi surface.

The extraction of the gap values from the conductance curves requires some kind of fitting to a suitable model. If one uses the single-band s-wave BTK model (even if generalized to the 3D case [37]), the fit of the curves is possible only in the region of the low-bias peaks, and completely fails at higher energies. Incidentally, it is interesting to observe that in this case, similarly to what reported in Ref. [23], many contacts in Sm-1111 apparently show a gap close to the BCS value. Instead, a fit of the *whole* conductance curves is possible *only* by using a two-band *s*-wave generalized BTK model. The agreement between the experimental curves and the theoretical ones is indeed rather good and the fit allows extracting two gap amplitudes, $\Delta_1$ and $\Delta_2$. In both La-1111 and Sm-1111 the two gaps are very different from each other, and their ratio $\Delta_2/\Delta_1$ is always around 3. The temperature dependence of these gaps is however different in the two compounds. While in Sm-1111 both the gaps approximately follow a BCS-like curve and clearly close at the same $T_c^A$, in La-1111 an anomalous suppression of both $\Delta_1$ and $\Delta_2$ occurs at $T \sim 0.8\, T_c^A$ (with $\Delta_2$ becoming undetectable at higher T and $\Delta_1$ showing a "tail" up to $T_c^A$). This anomaly, whose reason is not clear yet, may question the interpretation of $\Delta_2$ as a superconducting gap (and, indeed, the possibility that $\Delta_2$ is not a superconducting gap but an order parameter of different origin has been explored elsewhere [28]). However, in Sm-1111 the temperature dependency of $\Delta_2$ is really the one expected for a superconducting gap, and it is difficult to imagine a completely different nature of the superconducting state in these two materials. Moreover, the similar values of $\Delta_2/\Delta_1$ and of the ratios $2\Delta_1/k_BT_c$ and $2\Delta_2/k_BT_c$ in the two materials cannot be simply a coincidence and points towards a similar two-band superconductivity in both the La- and Sm- based compound. It should be borne in mind that, up to now, clear (but indirect) evidences of two-band superconductivity have been reported in literature in both these systems [20], [21], [22], and unambiguous, direct observations of two gaps such that $\Delta_2/\Delta_1 \approx 3$ have been reported for the 122 family [18]. These high values of the ratio $\Delta_2/\Delta_1$, here observed also in La-1111 and Sm-1111, certainly cannot be explained by the simplest interband-only, extended s±-wave model with only two bands with identical densities of states, where two order parameters with equal or very similar amplitudes are predicted [19]. The band structure of iron pnictides is however much more complicated, featuring at least two hole bands and two equivalent electron ones. As a matter of fact it can be shown that more complex interband-only, s±-wave models that take into account this three-band nature of the FS of Fe-based pnictides in the conventional BCS framework are able to reproduce the presence of two (or three) gaps with large $\Delta_2/\Delta_1$ ratios [43], [49]. However, this is not true in the Eliashberg formalism where a non-null intraband coupling of non-magnetic origin should also be taken into account to obtain two gaps with such different values.

Even if several experimental facts progressively seem to converge to a unified picture of superconductivity in Fe-based compounds, the final answer (at least for what concerns the number and amplitude of the gaps) will probably come from spectroscopic measurements in single crystals of 1111 compounds, as it happened for the 122 family. Unfortunately, present-day growth techniques do not allow the production of La-1111 single crystals large enough for this kind of measurements. As for Sm-1111, sub-mm crystals have been successfully grown and the PCAR spectroscopy measurements we carried out up to now perfectly confirm the findings shown here for polycrystals [50].

## Acknowledgements

The authors wish to thank L. Boeri, L.F. Cohen, R. De Renzi, O. Dolgov, Y. Fasano, S. Massidda, G. Sangiovanni, P. Szabó, and A. Toschi for useful discussions. Special thanks to I.I. Mazin for important comments and continuous encouragement. This work was partially supported by the PRIN project No. 2006021741. V.A.S. acknowledges support by the Russian Foundation for Basic Research Project No. 09-02-00205.